\documentclass[10pt,prl,twocolumn,preprintnumbers,amsmath,amssymb]{revtex4}
\usepackage[latin1]{inputenc}
\usepackage{graphicx}
\usepackage{wasysym}
\usepackage{color}

\begin{document}
\title{A little inflation in the early universe at the QCD phase transition}
\author{Tillmann Boeckel}
\affiliation{Institut f\"ur Theoretische Physik, Universit\"at Heidelberg\\ 
Philosphenweg 16, D-69120 Heidelberg, Germany}
\author{J\"urgen Schaffner-Bielich}
\affiliation{Institut f\"ur Theoretische Physik, Universit\"at Heidelberg\\ 
Philosphenweg 16, D-69120 Heidelberg, Germany}
\date{\today}

\begin{abstract} 
  We explore a scenario that allows for a strong first
  order phase-transition of QCD at non-negligible baryon number in the
  early universe and its possible cosmological observable
  consequences. The main assumption is a quasi-stable QCD-vacuum state
  that leads to a short period of inflation, consequently diluting the
  net baryon to photon ratio to it's today observed value. A strong
  mechanism for baryogenesis is needed to start out with a baryon
  asymmetry of order unity, e.g. as provided by Affleck-Dine
  baryogenesis. The cosmological implications are direct effects on
  primordial density fluctuations up to dark matter mass scales of
  $M_{max} \sim 1 - 10 M_{\astrosun}$, change in the spectral slope up
  to mass scales of $M_{max} \sim (10^6 - 10^8) M_{\astrosun}$,
  production of primordial magnetic fields with initial strength up to
  $B_{QCD} \sim 10^{12}$~G and a gravitational wave spectrum with
  present day peak strain amplitude of up to $h_c(\nu_{peak}) \sim
  5 \cdot 10^{-15}$ around $\nu_{peak} \sim 4 \cdot 10^{-8}
  \mbox{Hz}$. 
\end{abstract}

\maketitle

The theory of quantum chromodynamics (QCD) predicts a phase transition
from a quark-gluon plasma to a hadron gas in the early universe at a
critical temperature $T_{QCD} \approx$ 150-200 MeV
\cite{Aoki06b,Karsch07}. Only at low net baryon density lattice gauge
theory indicates a rapid crossover from the quark-gluon-plasma to the
hadronic phase. In the standard hot big bang
scenario the baryon asymmetry is $\eta_B \sim 10^{-9}-10^{-10}$
already before the QCD phase transition and therefore the idea of a
first order QCD phase transition in the early universe has been more
or less abandoned. However, most of the QCD phase diagram is actually
not well known. There has been recent progress in the attempt to
include a finite baryon density on the lattice \cite{Fodor04,Karsch02}
but in general one still has to rely on effective models \cite{PisarskiWilczek84} to tackle the QCD phase diagram. However, a true first order phase transition is expected at finite baryon
densities, as indicated by chiral effective models of QCD
\cite{StephanovRajagopalShuryak98} due to the melting of quark and/or
gluon condensates and the phenomenon of color superconductivity
\cite{Alford08}. Therefore, we would like to reopen the issue of a first order
cosmological phase transition by addressing whether there is a simple
scenario in which the QCD phase transition at finite baryon densities
can have consequences on cosmological scales. \\
In this letter we demonstrate that the scenario of a little inflation
at the QCD phase transition at high baryon densities is possible and not in
contradiction to present cosmological observations. It has interesting
cosmological implications though as it can directly affect primordial
density fluctuations on dark matter mass scales below $M_{max} \sim
1-10 M_{\astrosun}$, change the spectral slope up to mass scales of
$M_{max} \sim (10^6 - 10^8) M_{\astrosun}$ due to the change of the
global equation of state, produce primordial magnetic fields that may
be strong enough to seed the presently observed galactic and
extragalactic magnetic fields and produce a spectrum of gravitational
waves around a peak frequency of $4 \cdot 10^{-8}$ Hz that may be
observable via pulsar timing in the future \cite{Jenet06,Kramer04}.
Dark matter properties are also strongly affected as the annihilation
cross section for cold dark matter has to be up to nine orders of
magnitude lower to give the right amount of dark matter today, which
can be probed at the LHC by detecting the neutralino with an
unexpected low annihilation cross section, and thermal warm dark
matter masses can be of the order of MeV without exceeding the
decoupling degrees of freedom of the standard model. 
Such a cosmological phase transition would then
bear more resemblance to the situation in heavy ion collisions or even
the centre of neutron stars than to the standard QCD phase transition
in the hot big bang scenario. Hence, the upcoming FAIR facility
would actually be a probe for the physics of the early
universe in this scenario.\\
For a first order QCD phase transition in the early universe to be
possible a nonvanishing baryochemical potential $\mu_B$ is necessary
where $\mu_B/T \sim \mathcal{O}(1)$. The present day baryon asymmetry
$\eta_B = (n_B - n_{\bar{B}})/n_\gamma$ has been experimentally found
to be $5.9\cdot10^{-10} < \eta_B < 6.4\cdot10^{-10}$ at $98\%$
confidence by combining big bang nucleosynthesis, cosmic microwave
background and large scale structure results \cite{Steigman08}. The
number of baryons in a comoving volume is constant and can be
estimated to be $N_B \approx a_i^3 \mu_{Bi} T_i^2 \simeq a_f^3
\mu_{Bf} T_f^2$ where the index $i$ refers to the initial values when
the vacuum energy starts to dominate over the radiation energy and $f$
to the final values after reheating. Therefore the initial ratio of
the chemical potential to the temperature can be higher by $\frac{\mu_{Bi}}{T_i} \simeq \theta^3 \frac{\mu_{Bf}}{T_f} \left(\frac{T_f}{T_i}\right)^3$ with $\theta=a_f/a_i$. If the timescale for the decay of the false
vacuum is short compared to the Hubble time then $T_i\simeq T_f$ and
already for $\theta \sim 10^3$ (corresponding to a little inflationary
period with $N\approx 7$ e-foldings) the initial baryon asymmetry
$\eta_{Bi}$ and $\mu_i/T_i$ will be of order unity. Hence, the
evolution of the early universe could pass then through the first
order chiral phase transition of QCD.
A well established mechanism for generating a high baryon number in
the early universe is the Affleck-Dine baryogenesis
\cite{AffleckDine85}. 
The Affleck-Dine mechanism produces in most cases far too much baryon
number, thus either additional fields or more sophisticated coupling
terms have to be introduced to reduce the initial baryon number
production or it has to be reduced afterwards. For the latter case an
obvious possibility would be a large entropy release that dilutes the
baryon to photon ratio for example by an inflationary period
(se e.g.~ref.~\cite{Linde85}). Affleck-Dine baryogenesis can in fact
produce $\eta_B \sim \mathcal{O}(1)$, where this is probably an upper
bound \cite{Linde85}.\\
We note that the scenario proposed here has some similarities to
thermal inflation as discussed by Lyth and Stewart \cite{Lyth95} as
both are late time inflation periods in addition to ordinary inflation
with a length of only about 10 e-foldings and may thus help to to
resolve partly the moduli problem. In ref.~\cite{Borghini00} the
production of quark stars with masses of $10^{-2}-10 M_\odot$ was
proposed within a scenario similar to the one discussed here but at
small baryon densities and without addressing the key consequences of
such a second late time inflationary period. In \cite{Schwarz09} it
was recently proposed that a large lepton asymmetry could also result
in a first order QCD phase transition in the early universe.\\
An important issue of this approach is the stability of the barrier
between the false and the true vacuum in the effective potential up to
very low temperatures. This is indeed the case in chiral models of QCD
including gluonic degrees of freedom in the form of a dilaton field in
which case the barrier only vanishes in the $T\rightarrow 0$ limit
\cite{Campbell90}.
Csernai and Kapusta found in ref.~\cite{Csernai92} only small supercooling of about 1\% below the critical temperature using values of the surface tension of about $\sigma \sim 50 $MeV/fm$^2$. The nucleation rate $\Gamma$ depends exponentially on the surface tension as well as on the free energy difference between both phases and it's ratio to the Hubble parameter $\Gamma / H$ exhibits a maximum around $\sim T_c / 2$. If $\Gamma / H$ does not exceed one at this point the phase transition fails and we find that keeping the other parameters used in ref.~\cite{Csernai92} the surface tension must exceed 450 MeV/fm$^2 \sim 3.7 T_c^3$. However, the precise value of the surface tension at the QCD phase transition at high densities is not known and has been a matter of debate, see e.g.\: the discussion in
ref.~\cite{Voskresensky:2002hu} giving a possible range of 
$\sigma= 50$ to 150 MeV/fm$^2$ without excluding even smaller or
larger values. At very high densities calculations of
the surface tension in the first order phase transition between color 
superconducting phases and nuclear matter arrive at 
surface tensions of up to 300 MeV/fm$^2$ \cite{Alford01}.The value of the bag constant used by Csernai and Kapusta is at the upper end of values considered in the literature (i.e.  $\mathcal{B} =  \left(235\mbox{MeV}\right)^4$) and a reduction to the value found in the original paper of the MIT group by fits to hadron masses ($\mathcal{B} =  \left(145\mbox{MeV}\right)^4$, see ref.\:\cite{DeGrand75}) also reduces the surface tension needed for nucleation to fail to a value of 124 MeV/fm$^2$. This of course only covers the initial failure to nucleate, but in general $\mathcal{B}$ and $\sigma$ will both be temperature dependent. After some supercooling (e.g. 7 e-foldings at most) $\Gamma / H$ must exceed one for inflation to end and the phase transition to occur. This could either take place due to a strong drop in the surface tension or even due to a complete vanishing of the barrier between the two phases in the effective potential. In the latter case the surface tension goes to zero and a spinodal decomposition takes place. This has been studied e.g. in \cite{Jenkovszky90} for a bag like model.
Strong sensitivities of nucleation rates on the surface tension have been
also found for high-density matter as encountered in the interior of
neutron stars or in core-collapse supernovae \cite{Mintz:2009ay}
so that nucleation timescales can easily be in the range of 
$\mu$s to the age of the universe.\\
The equation of state has to fullfill the usual condition $\epsilon + 3 p < 0$ to enter an inflationary phase. In the Bag model this would be the case below a temperature $T_{inf}=\left(30 \mathcal{B}/(g\pi^2)\right)^{1/4}$. In the linear-$\sigma$-model or the NJL-model this occurs when the thermal contributions to the energy density become smaller than the vacuum contributions like the quark condensate $\left<m_q q \bar{q}\right> \approx f_\pi^2m_\pi^2$ and the gluon condensate $\beta_{QCD}/(2g) \left<G^a_{\mu\nu}G_a^{\mu\nu}\right> \approx 4\mathcal{B}$.\\
In ref.~\cite{Scavenius99} the idea of a "quench" in context
of heavy ion collisions is discussed, i.e.\ the chiral field is
trapped in a metastable minimum and supercools until the barrier in
the effective potential disappears at zero temperature and the field
"rolls down" to the true minimum. All in all a delayed chiral phase
transition at high baryon densities can not be excluded for the early
universe with our present poor knowledge of QCD at non-zero baryon
densities.\\
The majority of dark matter candidates is already chemically decoupled
from the radiation fluid at the QCD phase transition and thus do not
participate in the reheating at the end of the inflationary period.
Therefore the dark matter number density is diluted by the same factor
$\theta^3$ as the net baryon number. Normally the dark matter mass
enclosed inside the Hubble horizon is of the order of $10^{-9}
M_{\astrosun}$ at $T_{QCD}\sim 170$ MeV, so any influence on
perturbations inside dark matter would not have any consequences on
larger scales. An inflationary period at the QCD-phase transition can
change this drastically, since the amount of dark matter enclosed
inside the horizon must be larger by a factor $\theta^3$ initially to
give the right amount of dark matter today. For a short inflationary
period, as discussed here, there is an additional effect on
perturbations that have physical wavenumbers $k_{ph} \lesssim H$ at
the beginning of inflation. For general relativistic ideal fluid
density perturbations the system of dynamical equations is closed by
Einstein's $R^0_0$-equation that reads $(k_{ph}^2+\dot H)\alpha=4\pi G
(\delta \rho + 3\delta p)$ in uniform expansion gauge (see
e.g.~\cite{Schwarz99}) where $\delta \rho$ and $\delta p$ are the sum
of the density and pressure perturbations, respectively and $\alpha$
is the perturbation of the lapse. The time derivative of the Hubble
parameter is given via the second Friedmann equation $\dot H = -4 \pi
G (\epsilon + p) = -4 \pi G
(\frac{4}{3}\epsilon_{Ri}\left(\frac{a_i}{a}\right)^4 +
\epsilon_{Mi}\left(\frac{a_i}{a}\right)^3) \propto \left(\frac{a_i}{a}\right)^q$,
where the subscripts refer to matter and radiation with $q=3$ to 4,
respectively, and the index $i$ to the onset of inflation. Comparing
this to the first Friedmann equation one finds that $H^2 = \frac{8\pi
  G}{3}\left(\epsilon_V +
  \epsilon_{Ri}\left(\frac{a_i}{a}\right)^4 + 
  \epsilon_{Mi}\left(\frac{a_i}{a}\right)^3\right)$ meaning that the
two scales differ by $|\dot H/H^{2}|^{1/2} \simeq
\left(\frac{a_i}{a}\right)^{q/2}$ which would be irrelevant for a long
inflationary period (with more than 50 e-foldings) since $\dot
H^{-1/2}$ then corresponds to an unobservably large length scale.
Therefore, one can expect three spectral regimes, $(k_{ph}/H)_i >
a_f/a_i$ (always subhubble), $a_f/a_i > (k_{ph}/H)_i >
(a_i/a_f)^{q/2}$ (intermediate) and $(k_{ph}/H)_i < (a_i/a_f)^{q/2}$
(unaffected). Translating this to the highest affected mass scale
involved we estimate $M_{max}\sim 10^{-8}
M_{\astrosun}~(a_f/a_i)^{3q/2}$. Above this scale the spectrum of density
perturbations is given by the primordial spectrum of density
perturbations, e.g.~a nearly scale invariant spectrum. Note that we do
not make a statement about the detailed evolution of perturbations
above or below these two scales at this point, we only stress that a
cosmologically interesting mass scale appears for a short period of
inflation that could lead to observable consequences. For a fully
consistent treatment of perturbations one needs to take into account
the dynamics of the chiral phase transition in a detailed model and
try to estimate the effects of reheating on the amplitude of
perturbations.\\
For cold dark matter the dilution of the energy and number densities
leads to the possibility of a matter dominated phase before the
inflationary phase since the dark matter energy density after
reheating is basically fixed by the present day value. Consequently the dark matter
density before inflation is larger by the same factor $\theta^3$ as the baryon density.
For $\theta \gtrsim 10^3$ a matter dominated phase is present before the QCD phase
transition and QCD inflation is naturally limited to a length of
$\theta^{inf}_{max} = \left(\frac{\mathcal{B}}{\epsilon_{DM}(a_f)}\right)^{1/3}
\approx 900 \left(\frac{\mathcal{B}^{1/4}}{235\mbox{MeV}}\right)^{4/3}
\left(\frac{0.236}{\Omega_{DM0}}\right)^{1/3}$. The highest
affected dark matter mass scale would be then $M_{max}\sim (10^{6}- 10^{8})
M_{\astrosun}$. One can put a general upper limit on the amount of
entropy that is released by demanding that the initial baryon
asymmetry is at most of order one, implying that
$\theta^{B}_{max} \lesssim (1/\eta_B(a_E))^{1/3} \approx 1200$ (a complete
spectrum of primordial fluctuations would require $\theta
\gtrsim 10^{10}$).\\
We note that for non-relativistic decoupling of dark matter the weak
interaction cross section will no longer give the right amount of dark
matter today, the dark matter annihilation cross section has to be
much smaller, i.e.\ $\sigma^{annih}_{dm}\sim \theta^{-3}\sigma^{weak}$
as $\Omega_{DM}\propto 1/\sigma^{annih}_{dm}$ (we ignore logarithmic
dependencies on the dark matter mass). This gives the interesting
prospect that the little inflation can be probed by the LHC since the
discovery of a standard weakly interacting massive particle like the
neutralino would exclude the scenario.\\
For thermally decoupled ultra-relativistic particles the ordinary
temperature relation to the radiation background is changed after
inflation $T = T_{DM} \theta
\left(g^s_{eff}(T_{Dec})/g^s_{eff}(T)\right)^{1/3}$. Generalizing the mass limit found in \cite{Boeckel07} one arrives at
$m^{max}_{DM} \approx 51 \mbox{eV} \theta^3
\left(\frac{4}{g_{DM}}\right)\left(\frac{g^s_{eff}(T_{Dec})}{106.75}\right)
\left(\frac{\Omega^0_{DM}h^2}{0.116}\right)$. This allows for a much
higher mass of a thermal relic particle without the need for a large
number of additional effective decoupling degrees of freedom beyond
those of the standard model.\\
A vanishing speed of sound during a first order phase transition can
also lead to the formation of primordial black holes (PBH) for a small
fraction of Hubble volumes that are sufficiently overdense
\cite{Jedamzik97}. The mass spectrum of these PBH will be strongly
peaked around $1 M_\odot$ which corresponds to the total (not just the
dark matter) energy density inside the Hubble volume at the phase
transition. The abundance of PBH depends on the spectral index and
amplitude of the density fluctuation spectrum that can differ
significantly around the Hubble scale in the presented scenario as
discussed above. Lumps of quark matter or small quarks stars could be
also produced but only with $M \sim 10^{-9} M_\odot$ as we argue that
nucleation starts after the little inflationary epoch.\\
For a first order QCD phase transition with bubble nucleation there is
a well discussed mechanism for producing magnetic fields via bubble
collisions \cite{ChengOlinto94}. Since the baryon number is carried by
massless quarks and massive nucleons in the respective phases the
baryon number will tend to concentrate in the quark phase, at least
close to the phase boundary \cite{ChengOlinto94}. Because
of their finite masses the muon and the strange quark are already
slightly suppressed at the critical temperature $T_c$ which leads to a
charge dipole layer at the phase boundary. The
resulting net positive charge density is $\rho_C^+ = \beta e n_B$ 
with $\beta \sim 10^{-2}-10^{-3}$ for a small
$\eta_B$ and $\beta=0.2$ for our case. Using the estimates of
ref.\cite{ChengOlinto94} we arrive at magnetic fields of strength
$B_{QCD}=10^8-10^{10}$G for low baryon asymmetry although MHD
turbulence may readily amplify the initial fields to the equipartition
value $B_{eq}=\sqrt{8 \pi T^4 v_f^2}$ (\cite{Sigl97} and
refs.~therein), where $v_f$ is the fluid velocity. In the little
inflation scenario the initial value of the baryon contrast between
the two phases can be much higher since nucleons will be highly
suppressed at $T\sim 170\mbox{MeV}/\theta\sim 0.2$MeV, while for a
random walk the baryon diffusion length $r_{diff} \propto
1/\sqrt{n_B+n_{\bar{B}}}\sim 4 \mu m~ \theta^{3/2} \sim 10$~cm is
larger because $n_B$ and $n_{\bar{B}}$ are reduced by a factor of
$\theta^3$. Altogether one can expect that the magnetic field $B$ will
easily reach an equipartition value of $B_{eq}\approx 10^{12}$G,
where $v_f\sim 1$ since the released latent heat is much larger than
the thermal energy.

The presently observed galactic and extragalactic magnetic fields have
strength $B^{obs}_\lambda=0.1-1 \mu$G, but the required seed fields
for an effective galactic dynamo mechanism on scales of 0.1 Mpc are
strongly model and parameter dependent and vary over many orders of
magnitude $10^{-30} G \lesssim B^{seed}_\lambda \lesssim10^{-10}$G
(see \cite{Widrow02} and references therein). In ref.~\cite{Caprini02}
it was argued that for a causal production mechanism the spectrum of
the generated magnetic field must be very blue for uncorrelated
superhorizon scales, i.e.\ $B^2_\lambda\propto \lambda^{-n}$ with $n
\geq 2$. Therefore $B^{seed}_{\lambda}$ can be strongly limited by the
allowed additional energy density at BBN \cite{Cyburt05}. We find that
the produced initial field corresponds to $B^{seed}_{0.1 Mpc}
<10^{-22}$G which translates to a mean field at the QCD scale of at
most $B_{QCD} = 5\cdot 10^{13}$G which is consistent with the above
estimates. In \cite{Caprini09} it was found that an inverse cascade mechanism
due to a non-vanishing helicity of the primordial magnetic field
(as one can expect in the presented scenario)  is able
to successfully seed large scale magnetic fields fields at the QCD phase transition.

In a first order phase transition nucleation can produce gravitational
waves due to bubble collisions and hydrodynamic turbulence as found by
\cite{KosowskyTurnerWatkins92}. For a nucleation rate of $\Gamma
\propto \exp{(t/\tau)}$ the peak frequency of the spectrum corresponds
to a present day frequency of $\nu^B_{peak} \approx 4.0 \cdot 10^{-8}
\mbox{Hz} \left(\frac{0.1 H^{-1}}{\tau}\right)
\left(\frac{T^*}{150\mbox{MeV}}\right)
\left(\frac{g^{eff}}{50}\right)^{1/6}$, where $T^*$ is the reheating
temperature. With the above estimates one arrives at
a peak strain amplitude $h_c(\nu^B_{peak})= 4.7 \cdot 10^{-15}
\left(\frac{\tau}{0.1 H^{-1}}\right)^2
\left(\frac{150\mbox{MeV}}{T^*}\right)
\left(\frac{50}{g^{eff}}\right)^{1/3}$ due to bubble collisions. The
kinetic energy of the colliding bubbles is also partially converted to
turbulent bulk motion of the plasma stirring gravitational waves at a
slightly lower frequency $\nu^T_{peak} \simeq 0.3\: \nu^B_{peak}$ with
a higher peak amplitude $h_c(\nu^T_{peak})\simeq 2.1
h_c(\nu^B_{peak})$ for a strongly first order phase transition
\cite{Kahniashvili08}. The approximate shape of the strain amplitude
spectrum is then $h_c(\nu) \propto \nu^{1/2}$ for $\nu < H$
(uncorrelated white noise) and $h_c(\nu) \propto \nu^{-m}$ for $\nu >
\nu^B_{peak}$ where the spectral index $m$ is at most 2 but could
easily be close to 1 or even lower due to multi bubble collisions
\cite{Kamionkowski94}. Pulsar timing already limits nucleation with
the presently available data to $\tau / H^{-1} < 0.12$ which will
improve to $\tau / H^{-1} < 0.06$ for the full data of the Parkes Pulsar
Timing Array project \cite{Jenet06}. The planned Square Kilometer
Array will be about four orders of magnitude more sensitive in
$\Omega_{gw}(\nu)$ \cite{Kramer04} which corresponds to one order of
magnitude improvement for the bound on $\tau / H^{-1}$. Detection via
the Laser Interferometer Space Antenna (LISA) could also be possible
if the high frequency tail of the spectrum has a spectral index
$m\lesssim 1.4$ and $\tau / H^{-1}\gtrsim10^{-2}$.

We have here only briefly introduced the idea of a little inflation at
the QCD phase transition and sketched the differences from the
standard scenario for structure formation, dark matter properties,
magnetic fields and gravitational wave production. The main
assumptions are a high initial baryon asymmetry before the QCD phase
transition and the existence of a quasi stable QCD vacuum condensate
that dominates the energy budget for a short period. Especially the
impact on structure formation in this approach seems to be rather
interesting but requires a more thorough field theoretical approach.

We thank Rob Pisarski, Eduardo Fraga and Ruth Durrer for
useful discussions and Arthur Kosowsky for
important comments on the gravitation wave signal. This work was
supported by the German Research Foundation (DFG) within the framework
of the excellence initiative through the Heidelberg Graduate School of
Fundamental Physics (HGSFP) and through the Graduate Program for
Hadron and Ion Research (GP-HIR) by the Gesellschaft f\"ur
Schwerionenforschung (GSI), Darmstadt.

\end{document}